\documentstyle[12pt,axodraw]{article}
\newcommand{\NP}{Nucl. Phys. }
\newcommand{\PR}{Phys. Rev. }
\newcommand{\PRL}{Phys. Rev. Lett. }
\newcommand{\PL}{Phys. Lett. }

\begin{document}
\baselineskip=20pt

\pagenumbering{arabic}

\null\vspace{2.0cm}

\begin{center}
{\Large\sf $O(\alpha^2 G_F m^2_t)$ Contributions to 
$H\to\gamma\gamma$ }
\\[10pt]
\vspace{2.0 cm}

{$\underline{\rm Yi~Liao{}^a}$ and Xiaoyuan~Li${}^{\rm b}$}

\vspace{3.5ex}
{a \small Department of Modern Applied Physics, Tsinghua University,
Beijing 100084, P.R.China\\}

{b \small Institute of Theoretical Physics, Chinese Academy of Sciences,\\
P.O.Box 2735, Beijing 100080, P.R.China \\}

\vspace{5.0ex}
{\bf Abstract}
\end{center}

The rare decay $H\to \gamma\gamma$ is a promising detection channel 
for an intermediate mass Higgs boson. We compute its two-loop
$O(\alpha^2 G_F m_t^2)$ correction in the standard model and find
that the relative correction to the decay rate 
runs between $0.7\%$ and $0.5\%$ for $M_H=80-150$ GeV.
The analogous correction to the amplitude 
for $gg\to H$ is recovered as a special case. The generalization 
of our result to other models is also briefly indicated.

\begin{flushleft}
\end{flushleft}

\vspace{2cm}
\begin{center}
             submitted to {\it Phys. Lett. B}
\end{center}

\baselineskip=24pt
\newpage

The standard model ( SM ) of the electroweak interactions 
$\cite{SM}$ has proved very successful in the description 
of electroweak phenomenology. Yet, there is one particle, 
the Higgs boson, predicted by the SM that has 
evaded detection up to now. The Higgs boson is an essential 
ingredient of the SM. It provides mass for the $W$, $Z$ 
bosons and fermions through the Higgs mechanism. 
Its discovery will thus be crucial for confirmation
of the Higgs sector in the SM.

The search for the Higgs boson is difficult however.  
Theoretically this is basically because the relevant 
parameters such as masses, couplings 
are essentially free in the SM. Although there are 
theoretical considerations that can more or less 
constrain them our knowledge about these parameters is 
mainly from the failure search of experiments.
For example, experiments at LEP I and SLC have ruled out 
the Higgs mass range $M_H \le 63.5$ GeV at the $95\%$ 
confidence level. It is expected that LEP II will extend 
the range up to $80$ GeV. Beyond this we have to appeal 
to the next generation of hadron colliders.
The production mechanisms for the Higgs boson at hadron 
colliders have been extensively studied in the literature
$\cite{Higgs}$. Generally the gluon fusion mechanism
$\cite{Georgi}$ dominates for a Higgs mass up to $700$ GeV. 
Above this range the dominant mechanism is through 
$W$, $Z$ scattering subprocesses. The detection mechanisms 
for Higgs boson in the range $M_H \sim 140-800$ GeV
have also been widely studied ( see, for example, 
Refs.$\cite{Higgs}\cite{Kunszt}\cite{Marciano}$). 
Less extensively studied is the so-called intermediate 
mass Higgs boson in the region $M_H\sim 80-140$ GeV
$\cite{Wudka}$. It is difficult to detect a Higgs boson 
in this region because the dominant decay mode 
$H\to b \bar{b}$ is badly buried in the enormous QCD 
jet background$\cite{Baer}$ so that one has to use rare
decays. A favorite decay mode is $H\to\gamma\gamma$, 
though an excellent energy resolution is still required 
to discriminate signals from the background produced 
from $q\bar{q}(gg)\to\gamma\gamma$ and fake
$\gamma$'s from $\pi^0$ decays$\cite{Kunszt}$. 
Since the signal of $H\to\gamma\gamma$ is so small and 
may be overwhelmed by the background it is important to
predict its decay rate as precisely as possible.

At lowest order the amplitude for the decay $H\to\gamma\gamma$ 
receives contributions from charged fermion and $W$ boson 
loops$\cite{Ellis}$. Among fermions in the SM, only the top 
quark is of practical importance in fermion loops$\cite{Georgi}$. 
The QCD correction to the top loop was considered in
Ref. $\cite{Bij}$ and found to be well under control. 
The $O(\alpha^2 G_F m^2_t)$ correction from internal exchange 
of the Higgs boson in the top loop was computed in 
Ref. $\cite{Liao}$. In this paper, we will complete this 
by including contributions from unphysical Goldstone bosons 
as well. The gluon fusion process $gg\to H$ is quite similar 
to the decay $H\to \gamma\gamma$. The QCD correction to the 
fusion was shown to be very large $\cite{Djouadi}$ and 
this would make the $O(G_F m^2_t)$ corrections relatively 
more important for $H\to \gamma\gamma$ than for $gg\to H$.
The $O(\alpha^2_s G_F m^2_t)$ correction to the fusion 
has been computed by the use of the low energy 
theorem for the trace of the energy-momentum tensor
and found to be small$\cite{Gambino}$. Although the 
$O(\alpha^2_s G_F m^2_t)$ correction for $gg\to H$ may be 
reproduced from that for $H\to\gamma\gamma$ by assigning equal 
charges to the top and bottom quarks, the latter 
cannot be obtained simply from the former as shown below.

The lowest order contribution to the amplitude for 
$H\to\gamma\gamma$ is
\begin{equation}     
\begin{array}{l}
i{\cal A}^{1-{\rm loop}}_{t+{\rm w}}
=i{\cal O} \displaystyle{ [N_c Q^2_t F_t+(Q_t-Q_b)^2 F_W],} \\
{\cal O} =\displaystyle{\frac{\alpha}{2\pi v} 
               \epsilon^*_{\mu}(1)\epsilon^*_{\nu}(2)
                (g^{\mu\nu}k_1\cdot k_2-k^{\mu}_2 k^{\nu}_1),}
\end{array}
\end{equation}
where $k_{1,2}$ and $\epsilon(1,2)$ are the momenta and 
polarization vectors of the two photons, $Q_{t,b}$ are the 
electric charges of the top and bottom quarks, $N_c$ is 
the number of colors, and $v=2^{-1/4} G_F^{-1/2}=246~$GeV. 
$F_t$ and $F_W$ are fuctions of $\eta_t=4m^2_t/M^2_H$
and $\eta_{\rm w}=4M^2_W/M^2_H$ respectively, given in 
Refs. $\cite{Higgs}\cite{Georgi}\cite{Ellis}$.
To obtain the desired $O(\alpha^2 G_F m^2_t)$ correction, 
we work in the heavy top limit. This is a good approximation 
for an intermediate mass Higgs boson. In explicitly renormalizable 
$R_{\xi}$ gauges, the leading term in this limit is provided
by internal exchange of the Higgs and unphysical Goldstone
bosons with minimal number of contact interactions amongst 
themselves. This is because the introduction of a contact 
scalar vertex will bring in a factor of $M^2_H=2 k_1\cdot k_2$ 
that cannot be cancelled by the possible logarithmic infrared 
behaviour of diagrams in the limit $m^2_t \gg M^2_H$, and 
thus will not contribute to the leading term.
Furthermore, the non-leading terms are $\xi_{W,Z}$-dependent, 
and this dependence is cancelled only when contributions 
from $W$, $Z$ bosons are included.  With these considerations, 
the relevant interaction Lagrangian is 
\begin{equation}     
\begin{array}{l}
{\cal L}
=\displaystyle{-\frac{m_t}{v}H\bar{t}t 
 +i\frac{m_t}{v} \phi^0 \bar{t} \gamma_5 t
 +\sqrt{2}\frac{m_t}{v} (\phi^+ \bar{t}_R b_L 
                      +\phi^- \bar{b}_L t_R)
             +eA_{\mu} (Q_t \bar{t} \gamma^{\mu} t
                   +Q_b \bar{b} \gamma^{\mu} b)}  \\
~~~~~\displaystyle{
        +ie(Q_t-Q_b)A_{\mu} (\phi^- \partial^{\mu} \phi^+
                         -\phi^+ \partial^{\mu} \phi^-)
        +e^2 (Q_t-Q_b)^2 A_{\mu} A^{\mu} \phi^+ \phi^- }, 
\end{array}
\end{equation}
where $H$, $\phi^{0,\pm}$ are the Higgs and unphysical 
Goldstone bosons respectively, and $A_{\mu}$ the photon 
field. We have ignored the small bottom mass and 
quark mixing.

The two-loop diagrams that may contain the desired
correction are obtained by attaching in all possible ways 
the two photon lines onto the diagram shown in Fig. 1. 
They are shown in Fig. 2. These diagrams are classified 
into three groups. 
The first group corresponds to the insertion of the 
following one-loop elements in the one-loop 
top or bottom diagram for $H\to\gamma\gamma$: 
the top self-energy (denoted as $W$ ),
the $Ht\bar{t}$ or $Hb\bar{b}$ vertex ( $V_1$ ), and 
the $A_{\mu}t\bar{t}$ vertex ( $V_2$ and $V_3$ ). 
The second group consists of the insertion of the one-loop 
$H\phi^+\phi^-$ vertex in the one-loop $\phi^{\pm}$ diagrams 
for $H\to\gamma\gamma$ ( $V_4$ ). 
The third group is the remaining overlapping diagram ( $R$ ). 
Note that diagrams $V_{3,4}$ and $R$ are present only when 
the exchanged virtual scalars are the charged one, $\phi^{\pm}$. 
It is these diagrams 
that will produce the difference between the $O(G_F m^2_t)$
corrections to $H\to\gamma\gamma$ and $gg\to H$.

The contributions from the $H-,~\phi^0-$ and 
$\phi^{\pm}-$exchanged diagrams are respectively 
gauge invariant and can be calculated separately. 
In addition, parity is violated by the $\phi^{\pm}tb$ vertex 
so the $\phi^{\pm}-$exchanged diagrams will generally induce 
a new parity-violating Lorentz structure,
$O_{\mu\nu}=i\epsilon_{\mu\nu\rho\sigma} 
k^{\rho}_1 k^{\sigma}_2$.
Since the one-loop amplitude is symmetric
and $O_{\mu\nu}$ antisymmetric with respect to $\mu$ and $\nu$, 
the parity-violating term will not contribute to the total 
decay rate to the order considered here and may be safely ignored. 
In other words, there are no ambiguities associated with the 
$\gamma_5$ problem though we are computing two-loop diagrams.

Now we describe the renormalization procedures used to 
define the physical parameters and renormalization constants. 
We use the dimensional regularization to regulate ultraviolate 
divergences and work in the on-shell renormalization scheme. 
For our purpose, the mass of the unphysical Goldstone bosons 
can be set to zero from the very beginning, but with one caveat 
as explained below.

$1.$ Mass and wavefunction renormalization of the top 
( $W$ ) -- The counterterms for the top mass 
( $\delta m_t$ ) and wavefunction renormalization constant 
( $\delta Z_t$ ) are determined by requiring that $m_t$ be the 
pole of the one-loop corrected propagator of the top and that 
the residue of the propagator at its pole be unity.
Note that in the case of $\phi^{\pm}$ exchange there are
wavefunction renormalization constants for the left- and 
right-handed parts respectively, 
$Z_t^L=1+\delta Z_t^L,~~Z_t^R=1+\delta Z_t^R$.

$2.$ Renormalization of the $Ht\bar{t}$ and $Hb\bar{b}$ vertices
( $V_1$ ) -- In the SM, this is not independent but is related to
the renormalization of the $t,~b$ self-energy. The Feynman rule for
the counterterm of the $Ht\bar{t}$ vertex induced by 
$H-$ or $\phi^0-$exchange is 
$-i m_t/v (\delta m_t/m_t +\delta Z_t)$.
For $\phi^{\pm}$ exchange there is no counterterm 
for the $Hb\bar{b}$ vertex in the limit of zero bottom mass,
but there is a counterterm for the $Ht\bar{t}$ vertex,
$-i m_t/v (\delta m_t/m_t +1/2 (\delta Z_t^L+Z_t^R))$
though its bare one-loop diagram does not exist in this limit.
Finally there is a global counterterm for the $Ht\bar{t}$ vertex,
$(-i m_t/v) (\delta v/v)$ where 
$\delta v$ is the VEV counterterm of the Higgs field induced
by the heavy top.

$3.$ Renormalization of the $A_{\mu}t\bar{t}$ vertices ( $V_{2,3}$ )
-- The electric charge is defined as usual in the Thomson limit.
We have checked that the $U(1)_{\rm e.m.}$ Ward identity is 
satisfied especially in the case of $\phi^{\pm}-$exchange where
this is not self-evident.

$4.$ Renormalization of the $H \phi^+ \phi^-$ vertex ( $V_{4}$ )
-- The computation of diagrams $V_4$ requires special care.
Let us first discuss the renormalization of the vertex.
The counterterm for the vertex is 
$-i (M^2_H /v) (\delta Z_{\lambda}-\delta v/v)\mu^{\epsilon}$,
where $\delta Z_{\lambda}$ is the renormalization constant
for the scalar self-coupling $\lambda$,
$ \delta Z_{\lambda}=\delta Z_H+(1/M^2_H) 
     ~(\delta M^2_H-\delta M^2_{\phi^{\pm}})+2 \delta v/v$.
The mass and wavefunction renormalization of the Higgs boson
$( ~\delta M^2_H, ~\delta Z_H ~)$ is done as for the top 
so that there is no renormalization factor for the external 
Higgs boson. The counterterm
for the mass of $\phi^{\pm}~(~\delta M^2_{\phi^{\pm}}~)$ is 
determined by the condition of tadpole cancellation
at the one-loop level.
The masslessness of $\phi^{0,\pm}$ is then automatically 
preserved by the Goldstone theorem at the same level. 
Since the counterterm for $V_4$ is to be inserted into 
a finite one-loop amplitude and only terms up to $O(m_t^2)$ 
are required, we obtain
\begin{equation}
\displaystyle
-i\frac{M^2_H}{v} (\delta Z_{\lambda}-\delta v/v) \mu^{\epsilon}
=i \mu^{\epsilon} 8 N_c \frac{m^2_t }{v} (\frac{m_t}{4\pi v})^2
  [\Delta_{\epsilon}-\frac{7}{48}\frac{M^2_H}{m^2_t}],
~~\Delta_{\epsilon}=\Gamma(\epsilon)+\ln \frac{4\pi \mu^2}{m_t^2},
\end{equation}
where $\displaystyle{\frac{\delta v}{v}
=\frac{7}{6}N_c(\frac{m_t}{4\pi v})^2}$ has been inserted.

It is straightforward to calculate the contributions from $H-$ or 
$\phi^0$-exchanged diagrams, i.e., diagrams $W$ and $V_{1,2}$. 
The bare H-exchanged diagrams sum to a finite, gauge
invariant form. In the heavy top limit, it is given by
\begin{equation}
i{\cal A}^{2-\rm loop}_{H{\rm (bare)}}
=\displaystyle{i{\cal O}
                N_c (\frac{m_t}{4\pi v})^2 [-6 Q^2_t] .}
\end{equation}
The counterterm diagrams as a whole are also finite and gauge invariant,
\begin{equation}
\displaystyle{i{\cal A}^{2-\rm loop}_{H{\rm (c.t.)}}
=\delta m^{(H)}_t \frac{\partial}{\partial m_t}
  i{\cal A}^{1-{\rm loop}}_t
=i{\cal O} N_c (\frac{m_t}{4\pi v})^2 [4 Q^2_t] 
,}
\end{equation}
where $\delta m^{(H)}_t$ is the counterterm for the top mass 
induced by H-exchange, and  $i{\cal A}^{1-{\rm loop}}_t$
must be computed in $n$-dimensions, i.e.,
\begin{equation}
\displaystyle{F_t=-\frac{4}{3}\Gamma(1+\epsilon)
\mu^{\epsilon}(\frac{4\pi \mu^2}{m^2_t})^{\epsilon}.}
\end{equation}
Note that the wavefunction renormalization constant for the top 
is cancelled. Similarly, we obtain,
\begin{equation}
\begin{array}{l}
i{\cal A}^{2-\rm loop}_{\phi^0{\rm (bare)}}
=\displaystyle{i{\cal O}
       N_c (\frac{m_t}{4\pi v})^2 [\frac{14}{3} Q^2_t],}\\
\displaystyle{i{\cal A}^{2-\rm loop}_{\phi^0{\rm (c.t.)}}
=\delta m^{(\phi^0)}_t \frac{\partial}{\partial m_t}
  i{\cal A}^{1-{\rm loop}}_t
=i{\cal O} N_c (\frac{m_t}{4\pi v})^2 [-\frac{4}{3} Q^2_t]. 
}
\end{array}
\end{equation}

It is much more difficult to calculate the contributions from 
$\phi^{\pm}$-exchanged diagrams. Besides $W, ~V_{1,2,3}$
and $R$, they involve $V_4$ that are seemingly infrared 
divergent in the limit of massless $\phi^{\pm}$. 
Although they are actually not infrared divergent, 
individual diagrams do contain terms that are non-analytic 
in $p^2=2 k_1 \cdot k_2$ ( Higgs boson momentum squared ), 
like 
$\displaystyle g_{\mu\nu} \ln^n \frac{p^2}{m^2_t}~(~n=1,2~)$,
$\displaystyle g_{\mu\nu} \frac{p^2}{m^2_t}\ln\frac{p^2}{m^2_t}$,
$\displaystyle \frac{k_{2\mu} k_{1\nu}}{p^2}$,
$\displaystyle \frac{k_{2\mu} k_{1\nu}}{p^2}\ln\frac{p^2}{m^2_t}$
and 
$\displaystyle k_{2\mu} k_{1\nu}\ln\frac{p^2}{m^2_t}$.
It is a non-trivial check of our calculation that these terms are 
cancelled in the sum of two diagrams in $V_4$ so that they
will not spoil our low energy expansion in the heavy top limit.
The imaginary part is also cancelled in the sum. This is
reminiscent of the observation that the one-loop 
$\phi^{\pm}$-exchanged amplitude for $H\to\gamma\gamma$
does not contain an imaginary part in the limit of massless 
$\phi^{\pm}$. 
The sum of all bare $\phi^{\pm}$-exchanged diagrams and the
counterterms for $V_4$ is
\begin{equation}
i{\cal A}^{2-\rm loop}_{\phi^{\pm}{\rm (bare)}}
=\displaystyle{i{\cal O}
N_c (\frac{m_t}{4\pi v})^2 
[(4 Q_t Q_b-\frac{8}{3}Q^2_t)+5(Q_t-Q_b)^2],}
\end{equation}
where the second term is contributed by diagrams $V_4$ and
countermterms, and the first term is contributed by other 
bare diagrams. We note that diagrams $V_4$ actually contain 
an additional term proportional to 
$\displaystyle{(Q_t-Q_b)^2 g_{\mu\nu} 
[-1-4\Delta_{2\epsilon}-\frac{2}{3}\frac{p^2}{m^2_t}]}$,
which is exactly cancelled by other diagrams.
The remaining counterterms also sum to a gauge invariant
form,
\begin{equation}
i{\cal A}^{2-\rm loop}_{\phi^{\pm}{\rm (c.t.)}}
=\displaystyle{
 \delta m^{(\phi^{\pm})}_t \frac{\partial}{\partial m_t}
 i{\cal A}^{1-\rm loop}_t}
=\displaystyle{i{\cal O}N_c (\frac{m_t}{4\pi v})^2 
[\frac{4}{3}Q^2_t]}.
\end{equation}
Notice that the wavefunction renormalization constants 
$\delta Z_t^L$ and $\delta Z_t^R$ are again cancelled
in the sum. This is because the absence of a counterterm
for the $Hb\bar{b}$ vertex is compensated for by the
presence of a counterterm for the $Ht\bar{t}$ vertex.
The same counterterm also makes the $\delta m^{(\phi^{\pm})}_t$
part just as simple as in the case of $H$ or $\phi^0$
exchange.

Finally, there is a contribution from the counterterm $\delta v/v$
for the $H\bar{t}t$ vertex, 
\begin{equation}
i{\cal A}^{2-\rm loop}_{\delta v}
=\displaystyle{\frac{\delta v}{v}i{\cal A}^{1-\rm loop}_t
=i{\cal O}N_c (\frac{m_t}{4\pi v})^2 
[-\frac{14}{9}N_c Q_t^2].}
\end{equation}

Three different methods are employed to compute two loop diagrams.
$(1)$ After loop integration we expand Feynman parameter integrals
in the heavy top limit to obtain the $g_{\mu\nu}$ and 
$k_{2\mu}k_{1\nu}$ terms. $(2)$ We slightly modify the Hoogeveen's
method on expansion in the large mass limit$\cite{Hoogeveen}$.
We directly expand top propagators to the desired order without
shifting $\gamma$ matrices from denominators to numerators 
beforehand, and then use algebraic identities to further reduce
their products. This simplifies algebra considerably.
We use this to get the $k_{2\mu}k_{1\nu}$ terms.
$(3)$ After loop integration we use numerical analysis to approach 
the heavy top limit. Both $g_{\mu\nu}$ and $k_{2\mu}k_{1\nu}$ terms
are computed. In the case of H-exchange all three methods lead to 
an identical result. For $\phi^0$-exchanged diagrams we use the first
two methods and indeed obtain the same numbers. It is 
complicated to apply the method $(2)$ to $\phi^{\pm}$-exchanged 
diagrams due to 
the infrared behaviour associated with masslessness of the bottom
and $\phi^{\pm}$, so only the method $(1)$ is used.
But even so, we still have nontrivial checks as mentioned above:
cancellation of non-analytic terms in diagrams $V_4$, and 
cancellation of divergent, non-gauge-invariant terms 
proportional to 
$g_{\mu\nu}(Q_t-Q_b)^2$ between $V_4$ and other diagrams.
To summarize, the complete $O(\alpha^2 G_F m^2_t)$ contribution
to the decay amplitude is
\begin{equation}
i{\cal A}^{2-\rm loop}
=\displaystyle{i{\cal O}N_c (\frac{m_t}{4\pi v})^2 
[4 Q_t Q_b+5(Q_t-Q_b)^2-\frac{14}{9}N_c Q_t^2].}
\end{equation}
The amplitude for the fusion $gg\to H$ is recovered by setting 
$Q_t=Q_b$ and changing coupling factors appropriately,
\begin{equation}     
\begin{array}{l}
i{\cal A}(g_a~g_b\to H) \\
=\displaystyle{
\frac{i \alpha_s}{2\pi v} 
\epsilon_{\mu}(1)\epsilon_{\nu}(2)
(g^{\mu\nu}k_1\cdot k_2-k^{\mu}_2 k^{\nu}_1)
{\rm tr} (\frac{\lambda_a}{2}\frac{\lambda_b}{2})
(-\frac{4}{3})[1+(\frac{m_t}{4 \pi v})^2(-3+\frac{7}{6}N_c)]
}         \\
=\displaystyle{
\frac{i \alpha_s}{2\pi v} 
\epsilon_{\mu}(1)\epsilon_{\nu}(2)
(g^{\mu\nu}k_1\cdot k_2-k^{\mu}_2 k^{\nu}_1)
{\rm tr} (\frac{\lambda_a}{2}\frac{\lambda_b}{2})
(-\frac{4}{3})[1+\frac{\sqrt{2} G_F m^2_t}{32 \pi^2 }],
}
\end{array}
\end{equation}
which coincides with the result of Ref. $\cite{Gambino}$.
\footnote{We thank A. Djouadi for pointing out an error in our 
comments on Ref. $\cite{Gambino}$ in the original version 
of this paper.}

Including $O(\alpha^2 G_F m^2_t)$ and QCD corrections, the decay rate for 
$H\to\gamma\gamma$ is
\begin{equation}     
\displaystyle\Gamma=\Gamma_{\rm 1-loop}
[1+\frac{\sqrt{2}G_F m^2_t}{8\pi^2} B+\frac{2\alpha_s}{\pi}C],
\end{equation}
where $\Gamma_{\rm 1-loop}$ is the lowest one loop contribution,
$C$ was computed in Ref. $\cite{Bij}$ and
\begin{equation}     
\displaystyle {B=N_c
[4 Q_t Q_b+5(Q_t-Q_b)^2-\frac{14}{9}N_c Q_t^2]
/[N_c Q_t^2 F_t+(Q_t-Q_b)^2 F_W ].}
\end{equation}
For numerical analysis we use $m_t=176$ GeV, then the 
$O(\alpha^2 G_F m^2_t)$
correction to the decay rate runs between $0.7\%$ and $0.5\%$ 
for $M_H=80-150$ GeV, which is roughly one half of
the corresponding QCD correction$\cite{Bij}$
if $\alpha_s \sim 0.1$ is used.

The results reported here may be employed to incorporate
contributions from exchange of extra scalars in models with
an extended Higgs sector. For example, in the two Higgs doublet model
$\cite{2HD}$ the contributions from all of the four physical scalars
may be incorporated by multiplying appropriate factors coming 
from vertices. But it is suspect that the heavy top limit remains
to be a good approximation as these particles are generally
much heavier than the top quark.

We thank K.-T. Chao and Y.-P. Kuang for discussions.

\newpage
\begin{flushleft}
{\Large Figure Captions }
\end{flushleft}
\noindent
Fig. 1  The two-loop diagrams are obtained
by attaching in all possible ways the two 
photon lines onto the diagram shown here. Solid and dashed
lines represent fermion and scalar fields respectively.

\noindent
Fig. 2  The two-loop diagrams to be computed here. 
Wavy lines represent photon fields.

\newpage

\begin{center}
\begin{picture}(450,300)(0,0)
\SetOffset(100,80)\SetWidth{2.}

\CArc(55,47.5)(47.5,0,180)\CArc(55,47.5)(47.5,180,360)
\DashLine(7.5,47.5)(102.5,47.5){4.}
\DashLine(55,95)(55,125){4.}

\SetOffset(300,80)
\Text(55,75)[]{\large Fig. $1$}
\Text(55,55)[]{\large Phys. Lett. B}
\Text(55,35)[]{\large Yi Liao}

\end{picture}\\
\end{center}

\newpage

\begin{center}
\begin{picture}(450,600)(0,0)
\SetOffset(10,80)\SetWidth{2.}

\SetOffset(0,400)
\Line(0,0)(110,0)\Line(0,0)(55,95)\Line(110,0)(55,95)
\DashCArc(27.5,47.5)(27.5,-120,60){4.}
\DashLine(55,95)(55,125){4.}
\Photon(0,0)(0,-30){5}{3}\Photon(110,0)(110,-30){5}{3}
\Text(55,-30)[]{\large $W$}
\SetOffset(0,-400)

\SetOffset(150,400)
\Line(0,0)(110,0)\Line(0,0)(55,95)\Line(110,0)(55,95)
\DashCArc(55,0)(27.5,0,180){4.}
\DashLine(55,95)(55,125){4.}
\Photon(0,0)(0,-30){5}{3}\Photon(110,0)(110,-30){5}{3}
\Text(55,-30)[]{\large $W$}
\SetOffset(-150,-400)

\SetOffset(300,400)
\Line(0,0)(110,0)\Line(0,0)(55,95)\Line(110,0)(55,95)
\DashLine(27.5,47.5)(82.5,47.5){4.}
\DashLine(55,95)(55,125){4.}
\Photon(0,0)(0,-30){5}{3}\Photon(110,0)(110,-30){5}{3}
\Text(55,-30)[]{\large $V_1$}
\SetOffset(-300,-400)

\SetOffset(0,200)
\Line(0,0)(110,0)\Line(0,0)(55,95)\Line(110,0)(55,95)
\DashLine(55,0)(27.5,47.5){4.}
\DashLine(55,95)(55,125){4.}
\Photon(0,0)(0,-30){5}{3}\Photon(110,0)(110,-30){5}{3}
\Text(55,-30)[]{\large $V_2$}
\SetOffset(0,-200)

\SetOffset(150,200)
\Line(55,0)(110,0)\Line(110,0)(55,95)
\Line(55,95)(27.5,47.5)\Line(27.5,47.5)(55,0)
\DashLine(0,0)(55,0){4.}
\DashLine(0,0)(27.5,47.5){4.}
\DashLine(55,95)(55,125){4.}
\Photon(0,0)(0,-30){5}{3}\Photon(110,0)(110,-30){5}{3}
\Text(55,-30)[]{\large $V_3$}
\SetOffset(-150,-200)

\SetOffset(300,200)
\Line(27.5,47.5)(82.5,47.5)\Line(82.5,47.5)(55,95)
\Line(55,95)(27.5,47.5)
\DashLine(0,0)(27.5,47.5){4.}
\DashLine(0,0)(110,0){4.}
\DashLine(110,0)(82.5,47.5){4.}
\DashLine(55,95)(55,125){4.}
\Photon(0,0)(0,-30){5}{3}\Photon(110,0)(110,-30){5}{3}
\Text(55,-30)[]{\large $V_4 ({\rm I})$}
\SetOffset(-300,-200)

\SetOffset(0,0)
\Line(0,47.5)(110,47.5)\Line(110,47.5)(55,95)
\Line(55,95)(0,47.5)
\DashLine(55,0)(0,47.5){4.}
\DashLine(55,0)(110,47.5){4.}
\DashLine(55,95)(55,125){4.}
\Photon(55,0)(20,-30){-5}{3}\Photon(90,-30)(55,0){5}{3}
\Text(55,-30)[]{\large $V_4 ({\rm II})$}

\SetOffset(150,0)
\CArc(55,47.5)(47.5,0,180)\CArc(55,47.5)(47.5,180,360)
\DashLine(7.5,47.5)(102.5,47.5){4.}
\DashLine(55,95)(55,125){4.}
\Photon(55,47.5)(55,77.5){5}{3}\Photon(55,0)(55,-30){5}{3}
\Text(75,-30)[]{\large $R$}

\SetOffset(300,0)
\Text(55,75)[]{\large Fig. $2$}
\Text(55,55)[]{\large Phys. Lett. B}
\Text(55,35)[]{\large Yi Liao}

\end{picture}\\
\end{center}

\end{document}